\documentstyle[preprint,aps]{revtex}
\tighten
\newcommand{\fig}{{FIG} }

\newcommand{\LJ}{Lennard-Jones}

\begin{document}

\title{ Critical Temperature and Nonextensivity in Long-range Interacting Lennard-Jones-like 
Fluids }
\author{ Sergio Curilef$^{1,2}$ and Constantino Tsallis$^{2,3}$}
\address{$^{1}$ Departamento de F{\'\i}sica, Universidad Cat\'olica del Norte, Casilla 1280, Av. Angamos 0610, Antofagasta, Chile}
\address{ $^{2}$ Centro Brasileiro de Pesquisas F{\'\i}sicas, Rua Xavier Sigaud 150,
22290-180 Rio de Janeiro-RJ, Brazil}
\address{ $^{3}$ Niels Bohr Institute, Blegdamsvej 17, DK-2100 Copenhagen, Denmark}

\address{}
\address{\mbox{ }}
\address{\parbox{14cm}{\rm \mbox{ }\mbox{ }\mbox{ }
\noindent
 Molecular dynamic simulations for systems with $D=2,3$ Lennard-Jones-like interactions 
are studied. 
 In the model, we assume that, at long distances, the two-body attractive
 potential decays as $r^{-\alpha}$.
 Thermodynamic extensivity (nonextensivity) is observed for $\alpha > D$
 ($0\leq \alpha \leq D$). Particular attention is payed to the liquid-gas
 critical point located, in the temperature-pressure plane, at ($T_c,P_c$).
 ($T_c,P_c$) are, in the $N\rightarrow \infty$ limit ($N\equiv$ number
 of molecules), {\em finite} for $\alpha > D$ and {\em diverge}
 for $\alpha \leq D$
 (as $(\alpha - D)^{-1}$ for $\alpha/D \rightarrow 1 + 0$). However,
 the variables $T_c^* \equiv T_c/N^*$ and $P_c^* \equiv P_c/N^*$ with
 $N^* \equiv [N^{1-\alpha/D} -1]/[1-\alpha/D]$ remain  
{\em finite for all} $\alpha$. 
Thus, the extensive and nonextensive regions become
unified, as recently conjectured. These results should be useful for discussing gravitation and some
special fluids.
}}
\address{\mbox{ }}
\address{\parbox{14cm}{\rm PACS numbers:  05.70.Ce, 05.70.Fh, 64.90.+b}}
 
\maketitle

\makeatletter


\narrowtext

In recent years, much attention has been paid to physical systems with 
microscopic long-range  interactions. 
Nonextensive behavior and difficulties with the standard formulation
of thermodynamics are since long known for this class of
systems (see, for instance, \cite{BHiPPS85,PJuPRB52}).
More precisely, in order to have a well defined thermodynamic limit in 
the usual sense, a system 
must have a {\it finite} free energy per particle. In other words, an 
increase of $N$ (number of
particles) is expected to leave the free energy per particle
unaltered if $N >> 1$, i.e., the free energy $F_N$ should be 
asymptotically proportional 
to $N$. However, when the effective range of the 
interactions between particles  decays slowly enough with the distance, 
$F_N$ increases with a higher power of $N$, hence $F_N/N$ diverges when 
$N\rightarrow \infty$.
According to many text-book interpretations of thermodynamics, the concept 
itself of thermodynamic 
limit is ill-defined in such cases and the formalism becomes unphysical (see, 
for instance, \cite{BHiPPS85}).
In some sense (that will become transparent later on), this is a restricted interpretation of thermodynamics and a possible way out from this 
difficulty has
been recently proposed \cite{TsaFRA95} and illustrated through several examples 
\cite{PJuPRB52,JGrPLA217,SCaPRB54}. In the present effort, we shall address along these
lines a Lennard-Jones-like fluid, with special emphasis onto its gas-liquid phase transition and
the associated critical point. To do this we shall perform molecular dynamics for the two- and
three-dimensional cases.

In order to understand the thermodynamical behavior of systems with
long-range interacting particles, we consider a \LJ-like model with 
pair interactions characterized by a~$(12,\alpha)$ potential.
A variety of values for $\alpha$ can be associated with  standard 
interactions in models for fluids. For instance,  for $D=3$, $\alpha=6$ 
corresponds to the standard  \LJ~fluid, $\alpha=3$ essentially corresponds 
to dipole-dipole interactions in systems with dipole configurations such 
that the interaction is attractive, $\alpha=2$ corresponds to 
dipole-monopole 
interactions (such us those relevant in the tides), $\alpha=1$ mimics the 
gravitational and the (unscreened) Coulombic  interactions, and finally 
$\alpha = 0$ corresponds to a mean field approach. For arbitrary $D$, the 
choice $\alpha = D-2$ corresponds to the isotropic solutions of the $D$-dimensional Poisson equation. Also, there are some special situations in which neutral, glassy, small spheres can interact through long-range potentials with low, and not even integer, $\alpha$. Such cases might occur, 
through a Casimir-like effect due to the large fluctuations at the critical 
point of standard fluids, within which the little spheres are immersed 
\cite{TBuPRL74}. We might occasionally refer to such fluids as Casimir-like 
ones. For all these reasons, we shall focus on herein all the values of 
$\alpha$ in the interval $0 \leq \alpha \leq 6$. 

An interesting though preliminary discussion of  this class of fluids has been done by Grigera~\cite{JGrPLA217}. Our aim in the present work is to
provide a more detailed description which will hopefully give a suitable picture
of some general thermodynamic properties of systems with long-range (and not singular at the origin) interactions. 

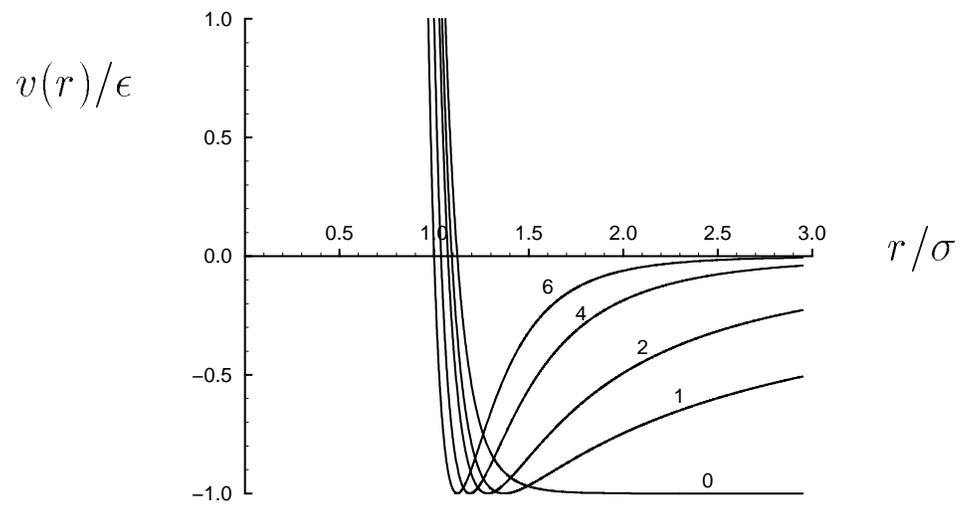
\begin{figure}
\caption{\LJ-like ~($12,\alpha$) potentials as functions of  $r$ for $\alpha = 0, 1, 2, 4 , 6$.}
\label{LJ1}
\end{figure}

The two-body potential we shall assume is given as follows:
\begin{equation}
v(r) = 4 \epsilon 
\left[ \left(\frac{\sigma }{r}\right)^{12}-  \frac{1}{4} \frac{(48/\alpha)^{\alpha/12} }
{(1-\alpha/12)^{1 - \alpha/12}}
\left(\frac{\sigma}{r}\right)^\alpha\right],
\label{LJP}
\end{equation}
where $r$ is the distance between particles and $\sigma >0$ and 
$\epsilon >0$ are specific \LJ-like parameters. 
For the long-tailed attractive term $r^{-\alpha}$ in potential~(\ref{LJP}) 
we consider $0 \leq \alpha < 12$. 
The  $r^{-12}$ term describes the
short range repulsive potential (due to nonbonding overlap between the
electron clouds for the $D=3$ case). 
The usual convention in the theory of fluids (for instance, standard \LJ~fluids) is 
to define the {\em diameter} $\sigma$, via $v(\sigma) = 0$.  Due to the range of the 
interactions, this would not seem particularly helpful here. So, we rather make a different convention, namely
the  $r^{-\alpha}$ coefficient has been chosen so that
the minimal value $v(r_{\mbox{min}}) = \epsilon$ in all cases. 
The \LJ-like $(12,\alpha)$ potentials as a function of $r$ are plotted in 
\fig\ref{LJ1} for typical values of $\alpha$.

A main difficulty for fluid systems with long-range (but not infinitely long) interactions is
the absence of any exact solution. In our present case,
calculations were performed with the molecular dynamics method
for systems of $N$ spherically-symmetric particles with periodic
boundary conditions.
Standard mean field (or van der Waals) theory for an {\it integrable} (i.e., $\alpha > D$)  
potential with a cutoff $r_c$ uses corrections $\propto  \int_{r_c}^\infty v(r)d^Dr$. However, our main aim is to discuss the trend of thermodynamic quantities as a function of $N$ 
(the size of the system) and to provide an unified approach to the thermodynamic limit for the entire range $(0 \leq \alpha \leq 6)$. 
Consequently, no corrections are considered in the present treatment and
a cutoff distance of half of the box size was applied to the
interaction; we verified that, in the $N\rightarrow \infty$
limit, no physical consequences seemed to emerge from the adoption of this computational convenience.
The equations of motion were solved using the Verlet algorithm~\cite{Allen-T}
and the
temperature was kept constant by a weak coupling of the system with
an external thermal bath~\cite{HBeJCP81}.

When starting a simulation from scratch, the initial configuration
is generated by randomly distributing the molecules. 
Then, the molecules
evolve in such a way as to achieve the
energetically most favorable positions and velocities. 
Full equilibrium is assumed only when the pressure, the potential
energy and the total energy exhibit stable values.
In general, the computational work is relatively heavy, since for systems 
with long-range interactions it is uneasy to obtain very precise numerical 
data for quantities such as the critical temperature, for example.  
Nevertheless, as we will see, the trends of the present numerical data clearly exhibit, in all the
cases we have studied here, the conjectured thermodynamical scalings \cite{TsaFRA95}. 

In the remainder of this paper, quantities are specified in usual reduced ({\it dimensionless})
form in terms of the \LJ$\;$parameters $\sigma$ and $\epsilon$.
The reduced number density is $\rho \sigma^D$ where
$\rho = N/V$; the reduced temperature is $ k_B T / \epsilon$
where $k_B$ is the Boltzmann constant; the reduced pressure is 
$\sigma^D P / \epsilon$; finally,
the reduced time is $\sqrt{\epsilon / m} \ \ t / \sigma$, where $m$ is the mass of the
particles. From now on we consider $\sigma=\epsilon=m=k_B=1$.

Special attention is payed to the point at which the 
{\em critical isotherm}  
($T=T_c$, where $T_c$ corresponds to the  {\em critical temperature}) simultaneous has a vanishing slope $\left( \partial P/\partial V \right)_{T_c} = 0$ and an inflection point, i.e., where the 
curve changes from convex to concave, hence, $\left( \partial^2 P/\partial V^2 \right)_{T_c} = 0$ in the pressure-volume plane. In fact, these properties simultaneously occur only at the {\em critical point} of the system. It is known that the standard \LJ$\;$fluid
provides\cite{PJePR220} a handy model for testing liquid theories and for investigating such phenomena as melting, the liquid-vapour surface, nucleation, etc. Consistently, our  interest
in the present more general fluid stems mainly in the critical point of the liquid-vapor curve.

The $NVT$-ensemble was used in our simulations. Naturally, strictly speaking, only at the thermodynamical limit we can speak of the critical temperature. Nevertheless, even for finite $N$, we can define an effective "critical" temperature $T_{c}(\alpha; N)$ by using the zero slope and inflection point criteria just mentioned. We have obtained the values for $T_{c}(\alpha; N)$ from simulations with systems containing $N = 27, 64, 125, 216$ and $512$ in three dimensions, and  
$N = 25, 49, 100, 225$ and $400$ in two dimensions.

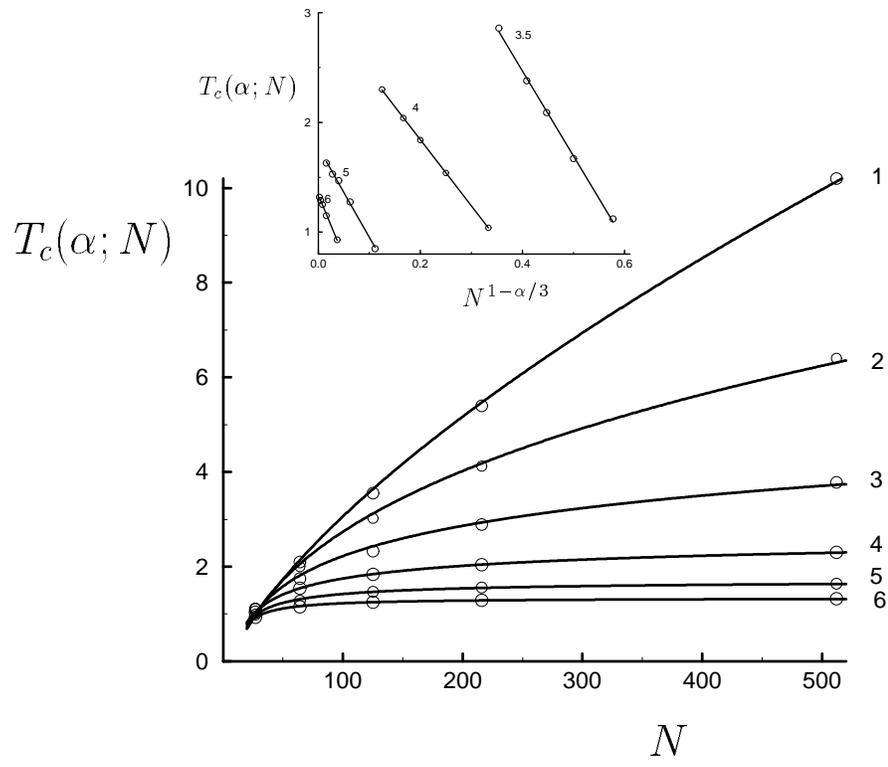
\begin{figure}
\caption{Approximate critical temperature $T_{c}(\alpha; N)$ 
as a function of $N$ for the $D=3$ system, for 
$\alpha = 1, 2, 3, 4, 5, 6$. Inset: In the extensive case ($\alpha > D$),
numerical data (circles)  were fitted using  
$T_c(\alpha;N)\approx T_c(\alpha) + a_\alpha N^{1-\alpha/D}$  (solid line),
for $\alpha = 3.5, 4, 5, 6$.}
\label{LJ2}
\end{figure}

In \fig~\ref{LJ2}, $T_c(\alpha;N)$ as a function of $N$ is depicted for 
typical values of $\alpha$ for D=3 case. 
For $\alpha > D$, the critical temperature 
shows a tendency  to saturate in the limit $N \rightarrow \infty$, 
but the convergence becomes very slow when 
$\alpha$ approaches $D$ from above, consistently with what was already
discussed for the total and the potential energies
of a similar fluid~\cite{JGrPLA217}; as we shall see, for $\alpha \le D$, 
the approximate critical temperature
diverges with $N$. A suitable plot for the curves associated with the 
extensive regime (i.e., $\alpha > D$) is depicted at inset in 
\fig\ref{LJ2}. 
Labels on lines correspond to several values of $\alpha$ at Figure
($\alpha =1,2,3,4,5,6$) and inset ($\alpha=3.5,4,5,6$).
The $D=2$ case exhibits similar trends.
In fact, all our results (for all $(\alpha,D)$) are satisfactorily 
fitted with the simple expression 
$T_{c}(\alpha; N) \approx T_c(\alpha) + a_\alpha N^{1-\alpha/D}$, 
where the parameters $T_c(\alpha)$ and $a_\alpha $ come out from the 
fitting. For the extensive region ($\alpha > D$)
$T_c(\alpha)$ represents the thermodynamic limit of the critical 
temperature; $a_\alpha < 0$. 

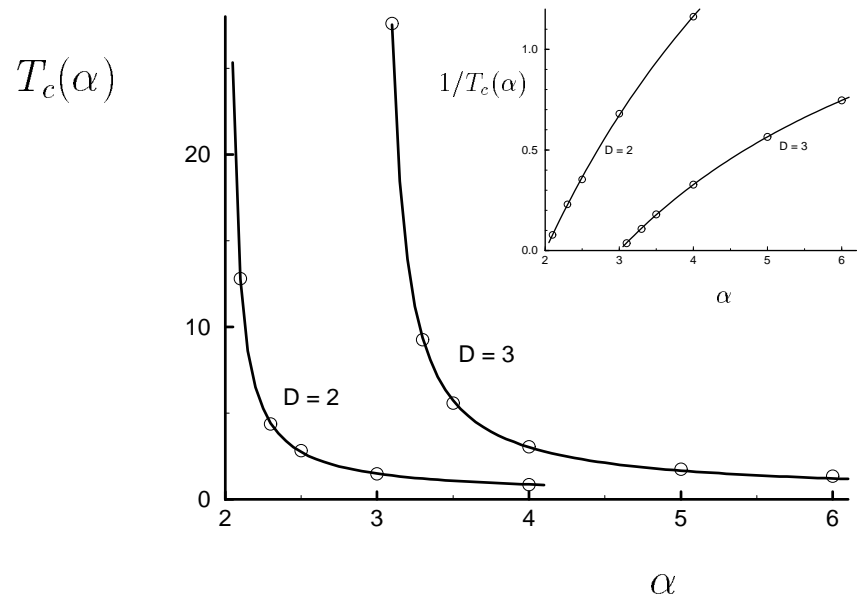
\begin{figure}
\caption{Thermodynamic limit critical temperature $T_c(\alpha)$ 
as a function of $\alpha$ for $D=2,3$. Inset: $1/T_c(\alpha)$ versus $\alpha$. For $\alpha \le D$
the critical temperature $T_c$ diverges.}
\label{LJ3}
\end{figure}

In \fig\ref{LJ3}, we depict some values of $T_c(\alpha)$ given by
the present approach as a function of $\alpha$. 
We verify that $T_c$ {\it diverges} for $\alpha \le D$ and that, when $\alpha \rightarrow D$ from above,
$T_c(\alpha)$ diverges like $ 1/ (\alpha-D)$ (see the inset in the figure). This behavior has been
conjectured \cite{TsaFRA95} for generic systems with long-range interactions.
The thermodynamic limit is not reached in the usual way in the nonextensive region. Nevertheless,
if we appropriately scale the various relevant quantities, our  
numerical results will show that this limit is in fact as well defined as that corresponding to the
extensive region. Let us now clarify how this happens. 

Thermodynamic quantities ${\cal A}_N$ like the internal energy, the free energy, Gibbs energy, 
etc., associated with systems including potentials having an attractive tail that decays as $r^{-\alpha}$, do not always scale with $N$. Indeed, we
generically have
\begin{equation}
\frac{{\cal A}_N}{N} \propto \int_1^{N^{1/D}} dr \frac{r^{D-1}}{r^\alpha} = 
\frac{1}{D} \frac{N^{1-\alpha/D} -1 }{1-\alpha/D},
\end{equation}
Consistently, we can introduce \cite{PJuPRB52,TsaFRA95} a new variable given by 
\begin{equation}
N^* \equiv \frac{N^{1-\alpha/D} -1 }{1-\alpha/D}
\label{Nestrela}
\end{equation}
which, in the limit $N \rightarrow \infty$, behaves asymptotically as
$1/(\alpha/D-1)$ for $\alpha/D > 1$; as $\ln{N}$ 
for $\alpha/D = 1$; and as $N^{1-\alpha/D}/(1-\alpha/D)$ 
for $0\leq\alpha/D < 1$.
For $\alpha=0$, $N^* \sim N$, which precisely recovers the value frequently used in mean field
approximations in order to renormalize the coupling constants in such a way as to make the system
to (artificially) become extensive. It is known since several decades (\cite{fisher} and references therein) that thermodynamical extensivity imposes, in classical systems like the present one,  $\alpha/D>1$. Our definition (3) clearly is consistent with this, but also provides the correct scaling for $0 \le \alpha/D \le 1$, where no essential knowledge has been developed until very recently, and along the present lines.

We expect the thermodynamic variables to generically include scaling with
$N^*$ as recently  conjectured \cite{TsaFRA95}. For instance, we expect the $\lim_{N \rightarrow
\infty} {\cal A}_N/(NN^*)$ to be finite {\it for all values} of $\alpha$, whether it is in
the extensive region or in the nonextensive one.
Thus, we can write
\begin{equation}
\frac{G_N}{N^*N} = \frac{U_N}{N^*N} - \frac{T}{N^*}\frac{S_N}{N} + 
\frac{P}{N^*}\frac{V_N}{N}.
\end{equation}

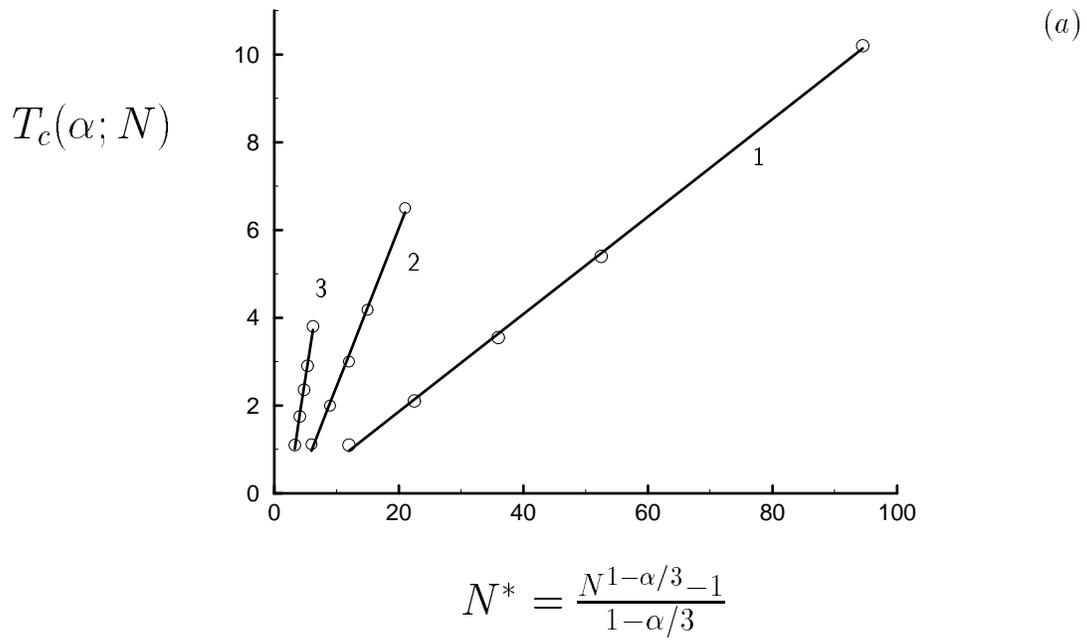
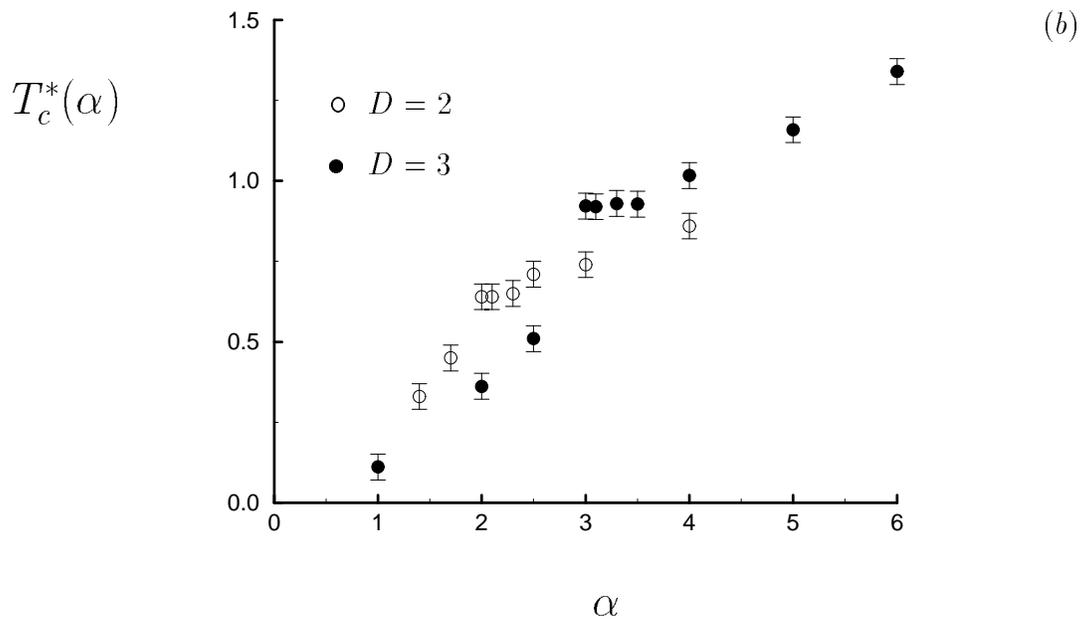
\begin{figure}
\caption{$(a)$ The $D=3$ approximate critical temperatures $T_{c}(\alpha; N)$ 
as functions of $N^*$ for $\alpha= 1, 2, 3$: Numerical data (circles)
and their fitting using $T_c(\alpha;N)\approx b_\alpha + T_c^*(\alpha) N^*$ (solid line); 
$(b)$ Thermodynamic limit for the scaled 
critical temperature $T_c^*(\alpha)$ as a function of $\alpha$ for the $D=2,3$ systems.}
\label{LJ4}
\end{figure}

It can be easily verified that this type of
scaling preserves the Legendre transformation structure of 
thermodynamics, even in the nonextensive region ($0 \leq \alpha \leq D$) \cite{TsaFRA95}.  
These features are in fact included in a more
general context, namely that of a recent statistical-mechanical formalism addressing
nonextensive systems~\cite{CTsJSP52,ECuJPA24,CTsCSF6}.
In what concerns the so called {\it intensive} variables ($T,P$, etc), we expect them to scale with
$N^*$. For instance, the critical temperatures must scale as 
$T_c(\alpha;N) \sim N^* T_c^*(\alpha)$. Indeed, in \fig\ref{LJ4}$(a)$, the critical temperature 
$T_{c}(\alpha; N)$ vs $N^*$ is exhibited
for typical values of $\alpha$, and a linear dependence with $N^*$
is clearly observed. 
As before, it is possible to
fit the data with the simple expression
$T_{c}(\alpha; N) \approx b_\alpha + T_c^*(\alpha) N^*$,
where $b_\alpha<0$ and $T_c^*(\alpha)$ are fitting parameters. Since $T_c^*$ is finite in all
cases, it is 
to be considered as the correctly scaled expression for the critical point of the system.
In \fig\ref{LJ4}$(b)$, we plot $T_c^*$ as
a function of $\alpha$ in both the extensive and nonextensive regimes. We see that
$T_c^*(\alpha)$ increases continuosly with $\alpha$ but its derivative 
$dT_c^*(\alpha)/d\alpha$ possibly presents a discontinuity at $\alpha = D$.
The error bars illustrate the mean error (less than 5\%) observed   
in the measurements of $T_c$ in our numerical simulations.

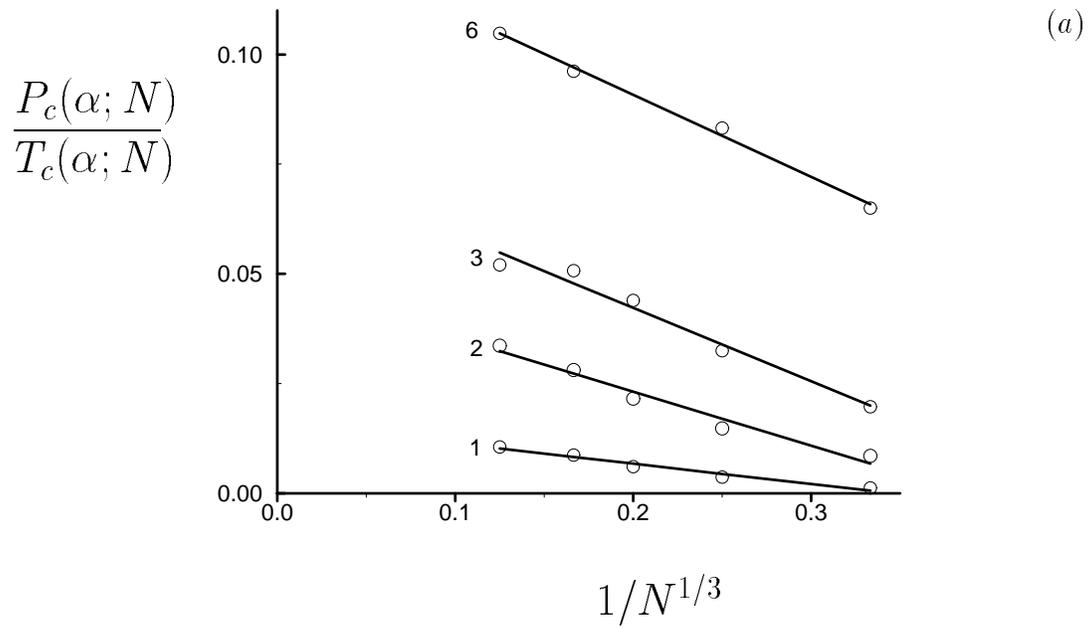
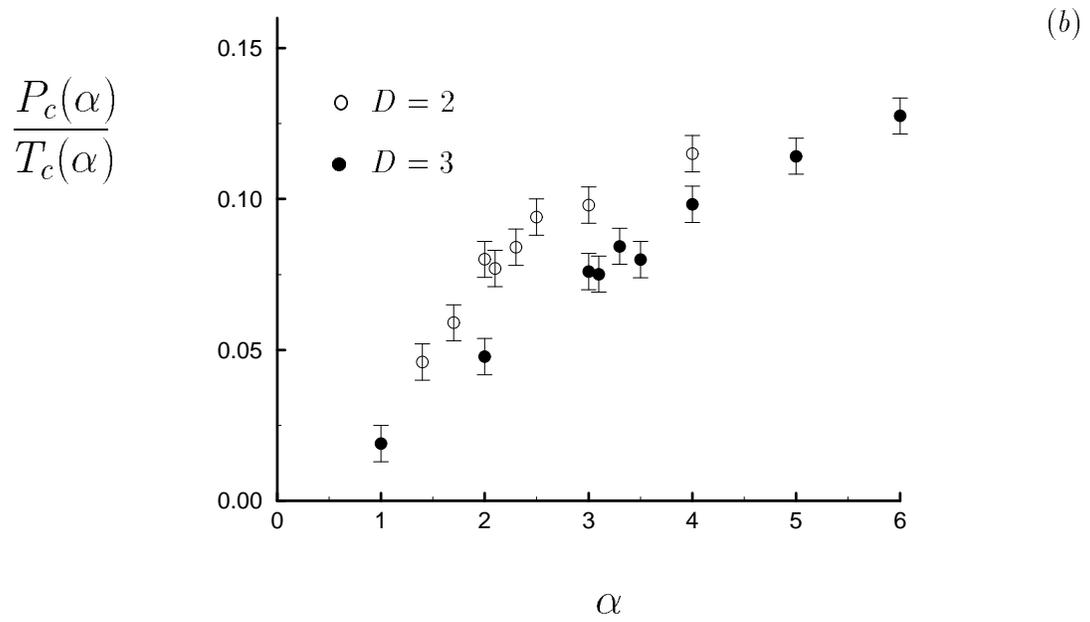
\begin{figure}
\caption{ $(a)$ The ratio $P_c(\alpha;N)/T_c(\alpha;N)$ 
versus $1/N^{1/3}$ in both extensive and nonextensive regions for the $D=3$ system; $(b)$ 
The thermodynamic limit values of $P_c(\alpha)/T_c(\alpha)$ versus $\alpha$ for the $D= 2, 3$ 
systems.}
\label{LJ5}
\end{figure}

In \fig\ref{LJ5}$(a)$, we plot  $P_c(\alpha;N)/ T_c(\alpha;N)$ as a 
function of $1/N^{1/3}$  for the $D=3$ system, and it turns out to be once again linear. This
enables a simple extrapolation (the precise finite size scalings associated with this problem have not been taken into account here, since they do not appear to add anything specially relevant for our present discussion) which provides 
$P_c(\alpha)/T_c(\alpha)$, which of course coincides with $P_c^*(\alpha)/T_c^*(\alpha)$, 
since both $P$ and $T$ scale with $N^*$.
We notice that, as it happened with $T_c^*(\alpha)$, the values are finite in both
the extensive and nonextensive regions (i.e., $\forall \alpha$): see \fig\ref{LJ5}$(b)$.
The observed error ($\approx$ 10 \% or less) in the measurement of the 
$P_c$ is roughly the double of that observed for $T_c$.

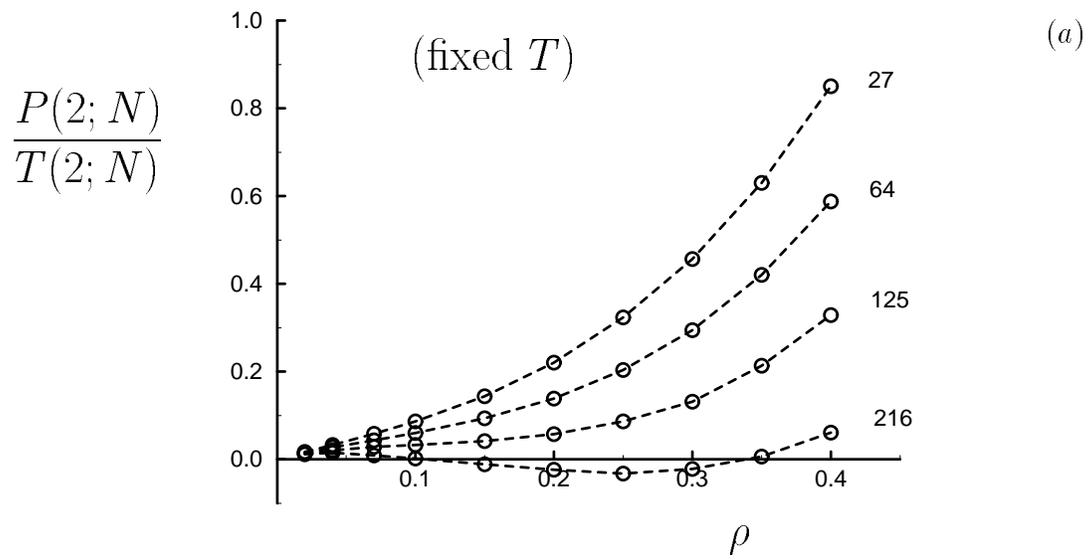
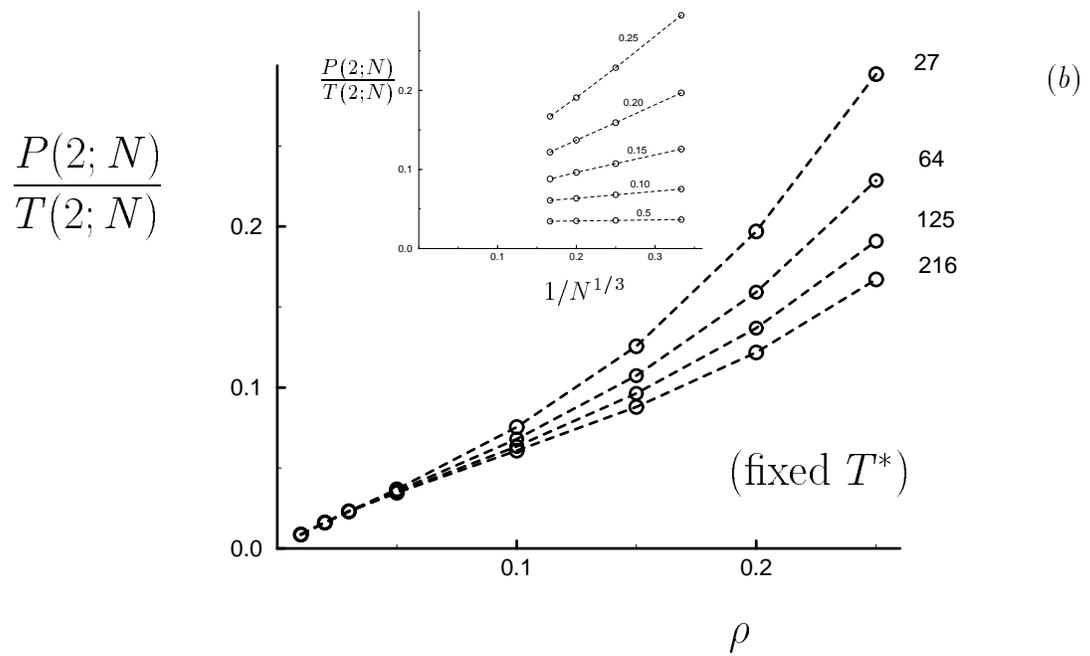
\begin{figure}
\caption{The ratio $P(\alpha)/T(\alpha)$ versus the density $\rho$ for $\alpha=2$ for the $D=3$ 
system with $N= 27, 64, 125, 216$ particles, at fixed $T=3.4$ $(a)$, and at fixed $T^* = 0.43$ 
$(b)$. 
Inset of $(b)$: The ratio as a function of $1/N^{1/3}$ for typical values of $\rho$ in order to 
illustrate their (linear) convergence.}
\label{LJ6}
\end{figure}

Let us finally address a more general situation, namely the functional relationships such as
the equation of states, in the present case the funtion $f(P^*,\rho, T^*) = 0$, which relates the
density $\rho$ with temperature and pressure. In \fig~\ref{LJ6}$(a)$, we illustrate the ideas with
the $D=3$ system using several sizes, namely $N = 27, 64, 125, 216$ at the same fixed temperature
$T=3.4$ and $\alpha = 2$. 
We notice, as expected, that the curves in the $(P,\rho)$ plane do not remain finite as $N$
increases. There is no usual thermodynamic 
limit and consequently no interesting properties can be exhibited under these conditions. 
The present class of models might produce a 
{\it negative} (here, unphysical) pressure $P_c(2;N)$ when
$N$ increases. Such behavior is commonly observed in $N$-dependent isotherms in the $P-\rho$ plane.
However, and interestingly enough, a completely different picture emerges if, instead of presenting the results at fixed $T$, we do it at fixed $T^*=T/N^*$. Indeed, in \fig~\ref{LJ6}$(b)$, we plot the curves corresponding to $T^* = 0.43$ ($> T_c^*(2)$) and we observe that the curves start converging as $N$ increases. In fact, the convergence can be exhibited versus
$1/N^{1/D}$ (see inset), which allows the establishment of a well defined 
thermodynamic limit even for this example in the nonextensive region. 
In addition to these results, 
we appoint that a new family of isotherms is obtained in the present
way, when 
the normalized function $T^*(\alpha; N)$ instead of $T(\alpha; N)$
is used, but  the true
temperature value is always given by $T(\alpha;N)$ in all cases. In
the same line, we verify that the values of 
$P(2;N)/T(2;N)$ appear to converge, as desired, to nonnegative 
values when $N \rightarrow \infty$. 

At this point, it is important to clarify that the thermodynamical variables we are analyzing
are those which appear in relations such as the equation of states. The parameters which
characterize thermodynamical equilibrium are the usual ones, i.e., the temperature $T$ of the 
external thermal bath, its pressure $P$, etc.

Summarizig, we have shown an example, a Lennard-Jones-like fluid, that can violate 
the usual scalings of thermodynamics, namely those
corresponding to extensive systems, i.e., those associated with interactions whose range is not too long,
more precisely satisfying $\alpha>D$. We have shown how the appropriate scaling (through $N^*$
given by definition (5)) of the various
thermodynamical quantities enables an unified picture for both short- and long-range interacting
systems. By so doing, we have empirically established a variety of secondary finite-size scalings.
In addition to this, we have verified that the critical temperatures, pressures, etc, 
diverge like $1/(\alpha - D)$ when
$\alpha$ approaches $D$ from above, a fact which is expected to be model-independent.
Let us finally add that we hope to have provided enough arguments to show that thermodynamics and the concept of
thermodynamical equilibrium can very well accomodate the (until now almost unexplored)
nonextensive systems. The detailed study of more such systems (quantum, frustrated, etc) 
certainly is very welcome.

One of us (S.C.) ackowledges partial support by FONDECYT, grant 3980014. 
The other one (C.T.) is thankful to Per Bak for useful remarks and warm 
hospitality at the Niels Bohr Institute where this work was concluded. 
Finally, CNPq and PRONEX (Brazilian Agencies) are also ackowledged for 
partial support.

\newpage

\end{document}